\documentclass[twocolumn,floats,prl,showpacs]{revtex4}

\usepackage{amsmath}
\usepackage{amsfonts}

\usepackage{graphicx}

\newcommand {\be}{\begin{equation}}
\newcommand {\ee} {\end{equation}}
\newcommand {\bea}{\begin{eqnarray}}
\newcommand {\eea} {\end{eqnarray}}
\newcommand{\non}{\nonumber}

\newcommand{\ket}[1]{\left| #1 \right>}
\newcommand{\bra}[1]{\left< #1 \right|}

\newcommand{\Hilb}{\mathcal{H}}

\begin{document}


\title{Real-space entanglement spectrum of quantum Hall systems}
\author{J. Dubail$^1$, N. Read$^1$, and E.H. Rezayi$^2$}
\affiliation{$^1$ Department of Physics, Yale
University, P.O. Box 208120, New Haven, Connecticut 06520-8120, USA\\
$^2$ Department of Physics, California State University, Los Angeles, California 90032, USA}

\date{March 12, 2012}

\begin{abstract}
We study the real-space entanglement spectrum for fractional quantum Hall systems, which maintains locality along the spatial cut, and provide evidence that it possesses a scaling property. We also consider the closely-related particle entanglement spectrum, and carry out the Schmidt decomposition of the Laughlin state analytically at large size.
\end{abstract}


\maketitle


It is by now widely accepted that topological phases of matter \cite{TQCreview} defy the paradigmatic
classification based on local order parameters and broken symmetries. Instead, quantum information concepts, such as quantum entanglement, have yielded valuable insight on this topic in recent years, and are believed to be good probes of topological order. For instance, the topological entanglement entropy \cite{KitaevPreskill,LevinWen} obtained from a spatial bipartition of the ground state of a fully-gapped Hamiltonian in two dimensions is one characteristic of a topological phase. At the end of Ref.\ \cite{KitaevPreskill}, Kitaev and Preskill (KP) showed that the same result can be obtained by assuming that the reduced density matrix has the form of the thermal density matrix of a gapless chiral energy spectrum, as would occur at an edge of a quantum Hall (QH) system.

To be precise, a bipartition of a quantum system is defined when the Hilbert space factors into two
parts, $\Hilb = \Hilb_A \otimes \Hilb_B$. Then the (normalized) ground state has a Schmidt decomposition \cite{nielsenchuang}:
\begin{equation}
	\label{eq:Schmidtdef}
	\ket{\psi} = \sum_{i} e^{-\xi_i/2}  \ket{\psi_{A,i}} \otimes \ket{\psi_{B,i}},
\end{equation}
where the $e^{- \xi_i/2}$'s ($e^{-\xi_i/2}>0$) are the Schmidt singular values, and $\ket{\psi_{A,i}}$ ($\ket{\psi_{B,i}}$) are an orthonormal set in $\Hilb_A$ ($\Hilb_B$). Equivalently, the reduced density matrix $\rho_A \equiv {\rm tr}_{\Hilb_B} \ket{\psi} \bra{\psi}$ has eigenvalues $e^{-\xi_i}$, and $\sum_{i} e^{- \xi_i}=\left<\psi | \psi \right>=1$. The set of $\xi_i$ is called the {\em entanglement spectrum} (ES).

For extended homogeneous systems defined as many-particle systems, lattice models, or field theories, with fairly short-range interactions, a natural way to partition is with a cut in position space (in continuous systems), or between a row or plane of sites (in a lattice model). It is for such types of partition that the results of Refs.\ \cite{KitaevPreskill,LevinWen} were derived. But studies of fractional QH states, in which the particles are confined to the lowest Landau level (LLL), have usually used other bipartitions. One of these is the {\em orbital partition} (OP) \cite{Schoutens1}, in which a basis set for the LLL single-particle states is divided into two subsets, and this leads to a bipartition of the $N$-particle Hilbert space also. If the basis set has the form that is natural in the Landau gauge (we will describe it for the plane here for simplicity, though later we use a compact geometry), then basis states are eigenstates of translations in the $y$ direction (with momentum eigenvalue $k_y$), and a Gaussian form in the $x$ coordinate (with center at $k_y$), so that partitioning into, say, the subsets $k_y<0$ and $k_y>0$ corresponds roughly to a spatial cut along the $y$ axis. For the LLL on a sphere, the corresponding scheme uses eigenstates of the $3$-component of angular momentum, and corresponds roughly to a spatial cut along a line of latitude.

Li and Haldane (LH) \cite{LiHaldane} studied the ES of QH states using OP on the sphere. They argued that the ES contains a low-lying part in which the multiplicities are related in a universal way to those in the chiral conformal field theory (CFT) which describes the edge physics of the topological phase. This low-lying part is usually well-separated from the rest of the ES, with a gap going to infinity for certain model wave functions such as the Laughlin state \cite{Laughlin}. LH proposed to use the low-lying part as a diagnostic tool for numerical calculations.

In this paper, we examine the direct {\em real-space partition} (RSP) in QH systems, defined in coordinate (not momentum) space, just as it is for non-QH systems. This bipartition by definition maintains locality along the cut in, say, the $y$ direction---that is, correlations (described by $\rho_A$) along the cut are short range. Such locality, which (as we will show) does not hold for the OP that involves instead a bound on $k_y$, is extremely helpful in theoretical analysis. First, for RSP of Laughlin states, we provide numerical evidence that a ``scaling form'' motivated by locality holds for the ES. Stated loosely, this says that as the thermodynamic limit is approached, not only the multiplicities but also the eigenvalues $\xi_i$ in the low-lying ES approach the energy spectrum of a local field theory on the cut, which is similar to the edge theory (for the trial states we study, this is a chiral CFT). This stronger form of the LH proposal resembles the idea of KP. It is not clear if the topological entanglement entropy \cite{KitaevPreskill,LevinWen} is obtained if this scaling form does not hold. Second, we also consider another bipartition for QH systems, called {\em particle partition} (PP) \cite{Schoutens1}, for which we extend the original definition, and show its close relation with RSP. Finally, we show for RSP and PP that the multiplicities for the Laughlin states obey the scaling form, and find the Schmidt decomposition explicitly for PP.

{\it Long-range correlations along the cut in OP---} We use the example of the equal-time Green's function (off-diagonal one-particle density matrix). In a large system of particles in the LLL in the $x$-$y$ plane, this takes a form independent of the details of the ground state other than the filling factor $\nu$, provided the state is translationally and rotationally invariant:
\be
\langle\psi|\hat{\psi}^\dagger(x',y')\hat{\psi}(x,y)|\psi\rangle
=\frac{\nu}{2\pi}e^{(z+\overline{z'})^2/4 - x^2/2-x'^2/2}
\ee
in the Landau gauge, where we use $z=x+iy$, $z'=x'+iy'$, and $\hat{\psi}(x,y)$ is the destruction field operator at $z$ (in this paragraph only, we set the magnetic length to $1$). This function falls off as a Gaussian in $|z-z'|$ in all directions. It can be derived by expanding $\hat{\psi}(x,y)$ in the (continuous) basis of LLL momentum eigenstates in the $y$ direction, which are $\propto e^{-(x-k_y)^2/2+ik_yy}$:
\bea
\lefteqn{\langle\psi|\hat{\psi}^\dagger(x',y')\hat{\psi}(x,y)
|\psi\rangle}&&\non\\
&=&\frac{\nu}{2\pi^{3/2}}\int_{-\infty}^\infty dk_y \, e^{k_y(z+\overline{z'})-k_y^2 -x^2/2-x'^2/2}.
\eea
For OP, we must restrict the integral to $k_y$ in part $A$, say $k_y<0$. Due to the discontinuity in the integrand of the Fourier transform, we find
\be
\frac{\nu}{2\pi^{3/2}}\int_{-\infty}^0 dk_y \, e^{k_y(z+\overline{z'})-k_y^2 -x^2/2-x'^2/2}
\sim \frac{\nu e^{-x^2/2-x'^2/2}}{2\pi^{3/2}i(y-y')}
\ee
as $|y-y'|\to\infty$; the long-range part has largest amplitude at $x=x'=0$, the ``cut''. We cannot evaluate other correlations as easily, but they will have related asymptotic forms. Thus in OP, correlations evaluated within part $A$ are long-ranged in coordinate space.

{\it Quantum Hall states on the sphere and RSP---} The single-particle states in the LLL on the sphere $S^2$ pierced by a rotationally-invariant magnetic field with a total of $N_\phi$ flux quanta ($N_\phi\geq 0$ an integer) \cite{HaldaneHierarchy} can be described using stereographic projection to the plane.
An orthonormal basis set of LLL states $\psi_m$ ($m=0$, $1$, \dots, $N_\phi$) with $3$-component of angular momentum $L_z = N_\phi/2-m$ on the sphere is thus given by $\psi_m\propto z^m e^{V(z)/2}$  up to a normalization factor (defined using the $L^2$ norm on the $z=x+iy$ plane), where the function $V(z)$,
\begin{equation}
	e^{V(z)} = \frac{1}{\left(1+ |z|^2 \right)^{2+N_\phi}},
\end{equation}
arises from the coordinate transformation from the sphere. The northern and southern hemispheres correspond to $|z|<1$ and $|z|>1$ respectively.

For a system composed of many identical particles, a bipartition can be defined by first dividing the single particle Hilbert space $\Hilb_{1}$ into two subspaces as a direct sum: $\Hilb_1=\Hilb_{1A}\oplus\Hilb_{1B}$. This then induces a corresponding bipartition of the $N$-particle space, in the form
\be
\Hilb_N=\bigoplus_{N_A=0}^N \Hilb_{N_A,A}\otimes \Hilb_{N_B,B},
\label{Hilb_part}
\ee
where $N_A$ ($N_B$) is the number of particles in part $A$ ($B$). For RSP, $\Hilb_1$ is an $L^2$ space of all normalizable functions over a manifold (the sphere in our case), and the two parts are simply the subspaces of normalizable functions supported on a subregion $A$ or on its complement $B$. Then any single-particle wavefunction can be decomposed into two orthogonal parts, each part lying in one subspace: $\psi({\bf r}) = \psi^A({\bf r}) + \psi^B({\bf r})$. The corresponding bipartition of the $N$-body Hilbert space is what we call RSP. In this paper, we usually take region $A$ to be the northern hemisphere, and $B$ the southern. We emphasize that for the QH problem, the $L^2$ space in which we decompose contains all the Landau levels and is infinite dimensional. Even though we consider ground states in the LLL, $\Hilb_A$ (and $\Hilb_B$) in the RSP decomposition include states that are not LLL states. The Schmidt decomposition is nonetheless of finite rank for a finite-size system.

\begin{figure}
	\begin{center}
	\includegraphics[width=0.4\textwidth]{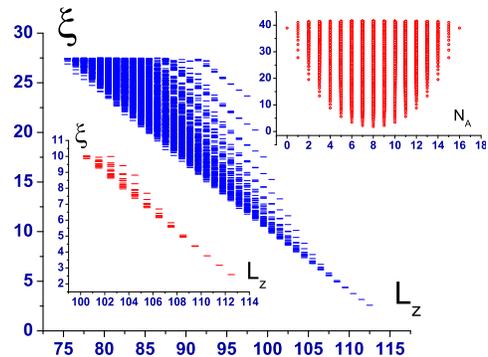}
	\end{center}
	\caption{(color online) Low-lying RSP ES of the $\nu=1$ filled LLL, for $N=30$ particles, and $N_A=15$. Lower inset: lowest levels. Upper inset: low-lying ES versus $N_A$ for $N=16$.}
	\label{fig:ESnu1}
\end{figure}

{\it RSP for the $\nu=1$ filled LLL--- } This case can be solved analytically as follows \cite{rstzv}. Let $c_m$
be fermion operators which destroy particles in the LLL basis states functions $\psi_m$. Then the filled LLL is $\prod_{m=0}^{N_\phi} c_m^\dagger|0\rangle$, where $|0\rangle$ is the vacuum with no particles.
For the Schmidt decomposition, we define operators $c_m^A$ and $c_m^B$ which destroy particles in the {\em normalized} states obtained from $\psi_m$ by multiplying it by $\Theta(1-|z|)$ or $\Theta(|z|-1)$ respectively, and normalizing. Then we have
\be
c_m = \alpha_m c_m^A+\beta_m c_m^B,
\label{alpha-beta_ops}
\ee
where $\alpha_m>0$, $\beta_m>0$ are the restricted norms:
\be
\alpha_m^2=\int d^2z |\psi_m|^2 \Theta(1-|z|),
\ee
and similarly for $\beta_m$ using region $B$. Substituting this into the ground state yields the Schmidt decomposition immediately. For each occupied basis state, the particle can be in either part $A$ or part $B$, and these occur with weights $\alpha_m$ or $\beta_m$. Hence each singular value $e^{-\xi_i/2}$ is given by a product of $\alpha_m$'s or $\beta_m$'s, one for each $m$. It follows that the terms $i$ in the Schmidt decomposition can be labeled by a set of ``occupation numbers'' $n_m=0$ or $1$ for each $m$, and the many-particle pseudoenergies $\xi_i$ are given by a constant plus a sum of single-particle pseudoenergies for the occupied modes: $\sum_{m=0}^{N_\phi} n_m  \varepsilon_m$ where
\be
\varepsilon_m = - \ln (\alpha_m^2 / \beta_m^2).
\ee
$\varepsilon_m$ is monotonically increasing with $m$, so the lowest $\xi$ is found by occupying all $m$ with $m\leq N_\phi/2$, and leaving the rest unoccupied (there is a zero mode $\varepsilon_{N_\phi/2}=0$ for $N_\phi$ even). The ES thus has exactly the form of the edge of the $\nu=1$ state. Here $N_A=\sum_m n_m$, $N_B=N-N_A$. A sample spectrum is plotted in Fig.\ \ref{fig:ESnu1}, for $N_A=N/2$ as a function of $L_z^A$, the angular momentum of part $A$. The linear dependence containing a velocity, which comes from the linearity of $\varepsilon_m$ near $m=0$, is evident. The full ES versus $N_A$ is also shown. The lower edge has a parabolic form, which we discuss below.

In performing the decomposition, we must reorder fermion operators, so that those referring to part $B$ are to the right of those for part $A$, because the tensor product in eq.\ (\ref{Hilb_part}) is ${\bf Z}_2$ graded. This introduces some minus signs, which become important for correlated states.

For the OP, the ES of the filled LLL contains only a single state, with $N_A$ and $L_z^A$ determined by filling all orbitals in part $A$. Thus the LH conjecture about the multiplicities does not hold in this case.

\begin{figure}
	\begin{center}
	\includegraphics[width=0.4\textwidth]{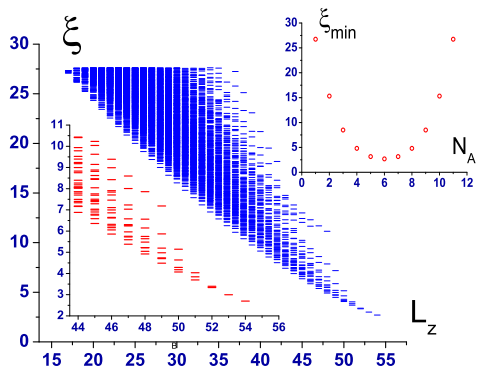}
	\end{center}
	\caption{(color online) Similar to Fig.\ \ref{fig:ESnu1}, but for the $\nu=1/3$ Laughlin state with $N=12$, $N_A=6$, and upper inset shows minimum $\xi$ only.}
	\label{fig:nu1_3}
\end{figure}

{\it Scaling conjecture---} Turning now to fractional QH states, we can state what we expect for the behavior of the ES in RSP. In the low-lying branch for generic ground states, or in the full ES for trial ground states, there is usually a single non-degenerate lowest pseudoenergy level $\xi_0$ at some values $N_{A0}$ and $L_{z0}^A$ of the good quantum numbers $N_A$ and $L_z^A$; these values depend on the system size $N$. Let us define quantum numbers and pseudoenergies $\Delta N_A$, $\Delta L_z^A$ and $\Delta \xi_i$ by subtracting off these values. We will say the ES has the {\em asymptotic scaling property} if for all $\Delta N_A$ and $\Delta L_z^A$, as $N\to\infty$, the set of $\Delta \xi_i$ approach the energy levels (minus the lowest energy) of a Hamiltonian that is the integral of a sum (with size-independent coefficients) of {\em local} operators in a 1+1-dimensional field theory on a circle. In particular, for certain trial states, such as the Laughlin and Moore-Read states, the field theory will be a chiral CFT that is the same one as for the edge theory. In the simplest case, the leading local operator allowed by symmetry will be the stress tensor $T(z)$, and then the low-lying spectrum will collapse onto a single straight line versus $\Delta L_z^A$ for each $\Delta N_A$. The coefficient is a ``velocity'' $v$ that actually has dimensions of length; if it is non-zero we have
\be
\Delta \xi_i=v L_0/R
\ee
where $L_0=-\Delta L_z^A+O(\Delta N_A)$ is the $i$th eigenvalue of the zeroth Virasoro generator, and $R$ is the radius of part $A$; in our scaling $v/R=O(N_A^{-1/2})$.
We emphasize that this scaling property encompasses both (i) the multiplicities and (ii) the pseudoenergies $\xi_i$ at each $\Delta N_A$ and $\Delta L_z^A$. LH's conjecture was that scaling holds for the multiplicities only. We conjecture that the scaling property holds in full for RSP ES of a gapped topological phase.

{\it Numerical results---} General trial states for a partially-filled LLL are linear combinations of Slater determinants, and the calculations resemble the steps above for the filled LLL, but the final Schmidt decomposition was performed numerically. Computations for RSP are much more costly than for OP of the same state. We present results for the Laughlin $\nu=1/3$ state in Fig.\ \ref{fig:nu1_3}. The ES is strikingly linear at small $\Delta L_z^A$, and multiplicities there are the same as in the edge of an infinite system for the lowest values, and the same as for OP over a larger range. We also show the lowest pseudoenergies as a function of $\Delta N_A$. The parabolic form should be compared with what is expected for the CFT of the edge theory (a chiral compactified scalar field \cite{WenEdge}), which by the scaling conjecture should be
\be
\Delta \xi_{\rm min}(\Delta N_A)=\frac{v}{R} \frac{Q(\Delta N_A)^2}{2},
\ee
for filling factor $\nu=1/Q$ (including $\nu=1$). For the data in Fig.\ \ref{fig:nu1_3}, this works quite well, by taking $v/R$ from the spectrum at fixed $\Delta N_A$. It gives a numerical estimate for the ``compactification radius'' for the scalar field, which should equal $Q$. Moreover, the general form of the spectrum is similar to that for $\nu=1$, for which we know analytically that the levels eventually collapse onto a straight line. While finite-size limitations have prevented us from making scaling plots to study the approach to scaling, it is at least plausible that it will hold based on the trends visible at these and smaller sizes.

\begin{figure}
	\begin{center}
	\includegraphics[width=0.4\textwidth]{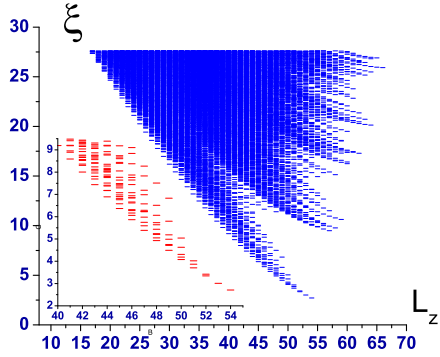}
	\end{center}
	\caption{(color online) Same as Fig. \ref{fig:nu1_3} but for the Coulomb interaction.}
	\label{fig:coulomb}
\end{figure}

For the LLL Coulomb interaction, the low-lying ES is similar to that of the Laughlin state---see Fig.\ \ref{fig:coulomb}.
We have also examined the ES for RSPs of the Laughlin states with cuts along different lines of latitude (not shown); these are always similar, but the minimum of the parabola of $\Delta \xi_{\rm min}$ versus $\Delta N_A$ is shifted.

{\it Particle Partition---} Another bipartition is the particle partition (PP) \cite{Schoutens1} in which one divides the $N$ particles into two sets of fixed sizes $N_A$ and $N_B$, and calculates the entanglement. The ES in PP has been considered recently \cite{BernevigPP}. We will first extend the definition so that it yields all values of $N_A$, as for OP and RSP. We introduce fictitious ``pseudospin'' variables for each particle; the two orthogonal pseudospin states will be called $|A\rangle$ and $|B\rangle$. To map a spinless ground state into the larger space that includes pseudospin, we assign to each particle $j$ the pseudospin state $(|A\rangle_j+|B\rangle_j)/\sqrt{2}$ (thus the state is still totally antisymmetric). We define a bipartition of this state in which the particles with pseudospin $|A\rangle$ constitute part $A$, and those with $|B\rangle$, part $B$. If operators $c_m^\sigma$ destroy particles in orbital $m$ and pseudospin $\sigma=A$, $B$, then the decomposition (\ref{alpha-beta_ops}) again applies ($c_m$ being for the superposition pseudospin state), but now with $\alpha_m=\beta_m=1/\sqrt{2}$ for all $m$. This is our extended PP; the part with any fixed $N_A$ is essentially the old PP. This scheme can be generalized by utilizing arbitrary values of $\alpha_m$, $\beta_m$ ($\alpha_m^2+\beta_m^2=1$); OP also fits into this by using values $\alpha_m=0$ or $1$. Now changing the set of $\alpha_m$ to another set of values (say, $1/\sqrt{2}$ for all $m$) defines a linear map $\Hilb_{N_A,A}\to \Hilb_{N_A,A}$ (more precisely, of the subspaces spanned by the basis states), which is invertible if all $\alpha_m$ are non-zero (and similarly for part $B$); it follows that the Schmidt rank in RSP is always the same as in PP, for each $N_A$ and $L_z^A$ (but generally larger than in OP).

\begin{figure}
	\begin{center}
		\includegraphics[width=0.4\textwidth]{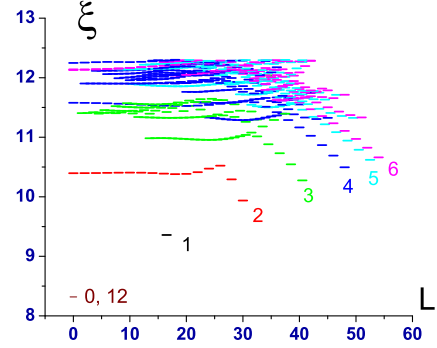}
	\end{center}
	\caption{(color online) Low-lying ES for the PP in the $\nu=1/3$ Laughlin state, for $N=12$ particles, including all $N_A$ values, versus $L^A$, not $L_z^A$.}
	\label{fig:PP}
\end{figure}

For PP, rotational invariance is preserved, so we may use the total angular momentum $L^A$ of part $A$ to label multiplets, each of which has degeneracy $2L^A+1$---see Fig.\ \ref{fig:PP}.
Unlike RSP, if $\xi_{\rm min}$ is plotted versus $N_A$, it yields an {\em inverted} parabola. For PP of the filled LLL (or for any single Slater determinant), $\xi_i=N\ln 2$ for all $i$.

{\it Explicit Schmidt decomposition of Laughlin states---} We perform the PP of the Laughlin state $\Psi_Q$ [with $N_\phi=Q(N-1)$] by letting the coordinates for part $A$ be $z_1$, \dots, $z_{N_A}$, and for part $B$, $w_1$, \dots, $w_{N_B}$. Then
the wavefunction is
\begin{eqnarray}
\label{eq:Laughlin_decomp}
\nonumber \Psi_Q &\propto&  \prod_{i,l} \left(z_i-w_l \right)^Q\prod_{i<j} \left(z_i-z_j \right)^Q  \prod_{k<l} \left(w_i-w_j \right)^Q \\
\nonumber 	&& \qquad {}\times e^{\sum_i V(z_i)/2 + \sum_l V(w_l)/2} \\
 &\propto& \exp \left( - \sum_{m > 0} j^A_{-m} j^B_{-m} \right) \, \Psi_Q(\{z\}) \widetilde{\Psi}_Q(\{w\}),
\end{eqnarray}
where we have used the following notations:
\begin{eqnarray}
\Psi_Q (\{z\}) &\propto& \prod_{1\leq i<j\leq N_A} \left( z_i-z_j \right)^Q e^{\sum_i V(z_i)/2}, \nonumber\\
\nonumber \widetilde{\Psi}_Q (\{w\}) &\propto& \prod_{k=1}^{N_B} w_k^{N_\phi} \prod_{1\leq i<j\leq N_B} \left(\frac{1}{w_i}-\frac{1}{w_j}\right)^Q e^{\sum_i V(w_i)/2}, \\
\nonumber j_{-m}^A &=& \sqrt{\frac{Q}{m}}\sum_{i=1}^{N_A} z_i^m,   \qquad j_{-m}^B = \sqrt{\frac{Q}{m}} \sum_{l=1}^{N_B} \left(\frac{1}{w_l}\right)^m,
\end{eqnarray}
and note $(z-w)^Q = (-w)^Q\exp(-Q\sum_{m=1}^\infty z^m w^{-m}/m)$ for $|z|<|w|$.
Note that $\Psi_Q(\{z\})$ is a droplet centered on the north pole, and $\widetilde{\Psi}_Q(\{w\})$ is centered on the south pole. Each factor $j^A_{-m}$ or $j^B_{-m}$'s creates a density excitation on the corresponding edge \cite{WenEdge}. The terms that are obtained by expanding the exponential in (\ref{eq:Laughlin_decomp}) are of the form
\be
\prod_{m >0} \frac{(j^A_{-m})^{n_m}}{\sqrt{n_m!}} \Psi_Q(\{ z \}),
 \ee
times similar for $B$. A linearly independent set for part $A$ is those with $n_m=0$ for $m>N_A$ and $\sum_m mn_m\leq QN_B$. Hence, this proves easily that in the limit $N_A$, $N_B\to\infty$, the count of levels for each $\Delta L_z^A$ and $\Delta N_A$ (defined using $N_{A0}=N/2$, $L_{z0}^A={\rm max}\,L_z^A$ at $N_{A0}$) matches that of the edge theory.

To complete the Schmidt decomposition, we need to know if these basis sets for parts $A$ and $B$ are orthonormal in the limit. This follows from existing results, in Ref.\ \cite{WenEdge} to leading order, and Ref.\ \cite{Wiegmann} to higher order. This implies that for PP, the $\Delta \xi_i$ in the scaling region obey the scaling property, with {\em vanishing} coefficient in the $O(N_A^{-1/2})$ term, that is vanishing velocity, for all $Q$ (this can also be shown for $\Delta N_A\neq0$). For $Q=1$, we know this holds to all orders. For all $Q$, if the scaling property holds, then the vanishing of the velocity follows from rotational invariance of PP. For RSP, we may imagine that a non-zero velocity appears by deforming the PP results due to the restriction on particle coordinates when calculating the overlaps. We present a detailed analysis elsewhere \cite{inprep}.

{\it Conclusion---} The RSP leads to very attractive results for the ES, and appears to obey the scaling property that results from locality. The PP is closely related, and allows an analytical treatment for the Laughlin state, in which the limiting ES in the scaling region has vanishing velocity.


\acknowledgments

This work was supported by a Yale Postdoctoral Prize Fellowship (JD), by NSF grant no.\ DMR-1005895 (NR), and by DOE grant no.\ DE-SC0002140 (EHR). After completing this work we became aware of similar work by others \cite{sterd}.


\begin{references}


\bibitem{TQCreview} For a review, see C. Nayak {\it et al.}, Rev. Mod. Phys. {\bf 80}, 1083 (2008).


\bibitem{KitaevPreskill} A. Kitaev and J. Preskill, Phys. Rev. Lett. {\bf 96}, 110404 (2006).

\bibitem{LevinWen} M. Levin and X.-G. Wen, Phys. Rev. Lett. {\bf 96},
110405 (2006).


\bibitem{nielsenchuang} M.A. Nielsen and I.L. Chuang, {\it Quantum Computation and Quantum Information} (Cambridge, 2000).

\bibitem{Schoutens1} M. Haque, O. Zozulya, and K. Schoutens, Phys. Rev. Lett. {\bf 98},
060401 (2007); O.S. Zozulya {\it et al.}, Phys. Rev. B {\bf 76}, 125310 (2007).

\bibitem{LiHaldane} H. Li and F.D.M. Haldane, Phys. Rev. Lett. {\bf 101}, 010504 (2008).

\bibitem{Laughlin}
   R.B. Laughlin, Phys.\ Rev.\ Lett.\ {\bf 50}, 1395 (1983).

\bibitem{HaldaneHierarchy} F.D.M. Haldane, Phys. Rev. Lett. {\bf 51}, 605 (1983).


\bibitem{rstzv} I. Rodriguez and G. Sierra, Phys. Rev. B {\bf 80}, 153303 (2009); A.M. Turner, Y. Zhang, and A. Vishwanath, Phys. Rev. B {\bf 82}, 241102 (2010).

\bibitem{WenEdge} X.-G. Wen, Int. J. Mod. Phys. B {\bf 6}, 1711 (1992).


\bibitem{BernevigPP} A. Sterdyniak, N. Regnault, and B.A. Bernevig, Phys. Rev. Lett. {\bf 106}, 100405 (2011).




\bibitem{Wiegmann} A. Zabrodin and P.B. Wiegmann, J. Phys. A {\bf 39}, 8933 (2006); P.B. Wiegmann, {\it private communication}.

\bibitem{inprep} J. Dubail, N. Read, and E. Rezayi, {\it in preparation}.

\bibitem{sterd} A. Sterdyniak {\it et al.}, Phys. Rev. B {\bf 85}, 125308 (2012).



\end{references}
\end{document}